\documentclass[preprint,showpacs,preprintnumbers,amsmath,amssymb]{revtex4}

\usepackage{graphicx}
\usepackage{dcolumn}
\usepackage{bm}

\begin{document}


\title{Studies of the $^{1}$S$_{0}$--$^3$P$_0$ transition in atomic ytterbium for optical clocks and qubit arrays}

\author{Tao Hong, Claire Cramer, Eryn Cook, Warren Nagourney, E. N. Fortson}

\address{Department of Physics, University of Washington, Seattle, WA 98195-1560}


\begin{abstract} 
We report an observation of the weak $6^{1}$S$_{0}$-$6^3$P$_0$ transition in $^{171,173}$Yb as an important step to establish Yb as a primary candidate for future optical frequency standards, and to open up a new approach for qubits using the $^{1}$S$_{0}$ and $^3$P$_0$ states of Yb atoms in an optical lattice. 
\end{abstract} 


\maketitle

\noindent Optical clocks show great promise as the next generation of precision frequency standards due to the high frequency and correspondingly high line-Q of weakly allowed atomic transitions.  Thanks to femtosecond comb technology \cite{Stenger:02}, clocks based on optical transitions in both single trapped ions and neutral atom clouds are poised to exceed the already very high precision set by microwave frequency standards \cite{Microwave}.  A recent proposal combining the Lamb-Dicke confinement of single ion frequency standards with the high signal-to-noise of neutral atom clouds, and exploiting a highly forbidden transition insensitive to external magnetic fields and trap light polarization has the potential to do even better \cite{Katori:01}.  The scheme makes use of the $^{1}$S$_{0}$-$^3$P$_0$ electric dipole transition in alkaline-earth atoms, weakly allowed in odd isotopes by the hyperfine interaction of the nuclear spin and completely forbidden by a single-photon process in even isotopes.  Trapping the atoms in a Stark-free optical lattice will allow a large ensemble of atoms to exhibit a spectrum free of Doppler and recoil shifts, leading to an estimated accuracy of a few parts in 10$^{18}$ \cite{Katori:03,Porsev:04}. As an additional application, the near absence of decoherence between the $^{1}$S$_{0}$ and $^3$P$_0$ states suggests forming a qubit from these states and creating an optical lattice array of such qubits for quantum information \cite{Knight:03,Duan:03}.

Yb and Sr are both strong candidates for an optical lattice clock as well as for creating useful qubits.  Both atoms have experimentally accessible optical transitions, allowing the atoms to be readily laser-cooled and trapped, and in the case of Yb, Bose-Einstein condensed \cite{YbBEC:03}.  While the Sr clock transition has been studied extensively, both in and out of an optical lattice \cite{French,Katori}, the same transition in Yb has thus far remained unexplored experimentally. In this paper, we focus on the odd isotopes of Yb \cite{Hong:05}.  Here we report the direct excitation of the 578 nm {$^{1}$S$_{0}-^{3}$P$_{0}$} transition in the odd isotopes $^{171,173}$Yb \cite{NISTYB:05}. We also discuss the possible application of this type of transition in making qubits for quantum computation.  

\begin{figure}[t]
\includegraphics[height=5cm]{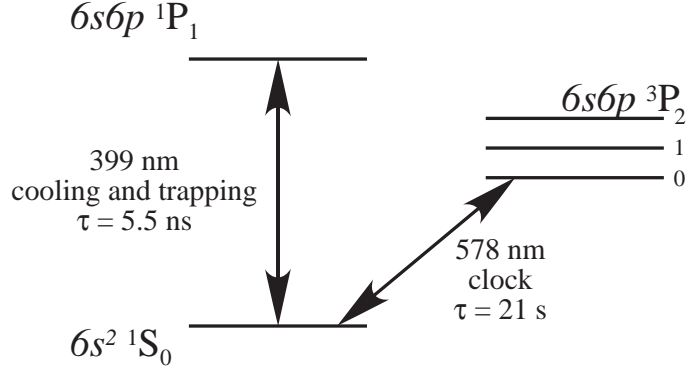}
\caption{ Yb energy levels, transition wavelengths, and lifetimes.}
\label{fig:levels1}
\end{figure}

The level structure of Yb is shown in Fig.~\ref{fig:levels1}.  The 
experiment is performed on a sample of cold atoms collected in a 
magneto-optical trap (MOT) operating on the strong 
{$^{1}$S$_{0}-^{1}$P$_{1}$} transition at 399 nm. To excite the clock 
transition, we irradiate the cloud of atoms with a 578 nm probe, 
continuously monitoring 399 nm fluorescence from the atom cloud.  As the 
probe laser frequency is scanned over the clock transition, there is an 
observable decrease in MOT fluorescence as atoms are excited into the 
metastable $^{3}P_{0}$ state and escape the trap.

\begin{figure*}[t]
\includegraphics[height=8.6cm,angle=270]{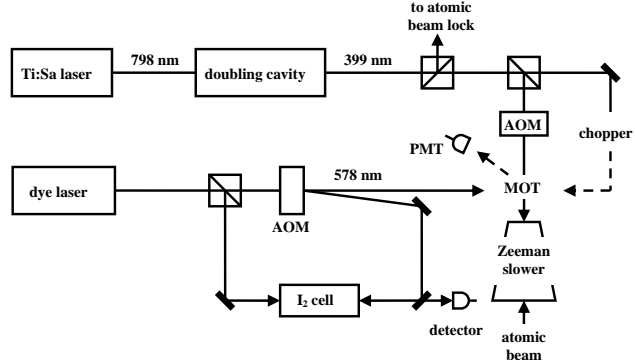}
\caption{Schematic of experimental setup:  A 
frequency-doubled Ti:Sapphire laser is used for the MOT and Zeeman slower. 
The Zeeman slower beam is detuned 300 MHz below the atomic resonance using 
a double-pass AOM.  Its 
frequency scan is calibrated with an I$_{2}$ saturated absorption cell. The blue MOT fluorescence is detected by a 
photomultiplier tube (PMT).}
\label{fig:apparatus}
\end{figure*}

The experimental apparatus is diagrammed in Fig.~\ref{fig:apparatus}.  The 
399 nm transition is used to decelerate an atomic beam in a Zeeman slower 
and then capture either $^{171}$Yb or $^{173}$Yb in the MOT.  The 399 nm 
output of a frequency-doubled Titanium:Sapphire laser (Coherent 899-21) is 
locked to the atomic resonance with an adjustable detuning to prevent slow 
frequency drift.  Further details of the MOT and Zeeman slower are 
discussed elsewhere \cite{Maruyama:03}.
We probe the clock transition with a ring dye laser (Coherent 899-21) internally locked to a temperature stabilized Fabry-Perot cavity.  A small fraction of the 
laser's output is directed through an acousto-optic modulator (AOM) to an 
iodine saturated-absorption cell serving as a frequency reference.  The 
main dye laser output is directed onto the MOT and its frequency 
continuously scanned from iodine line 1852 through the Yb clock resonance. 
The MOT beams are chopped on and off at a rate of 2.5 kHz.  Resonance data
are taken either with the probe beam on continuously or with the probe and
MOT beams chopped out of phase with each other. The latter procedure
should completely avoid the shifts and broadening of the clock transition
due to near-resonant 399 nm light while maintaining a steady-state MOT
population.
 The MOT magnetic fields produce a negligible perturbation on the clock resonance because of the small g-factors of the $^{1}$S$_{0}$ and 
$^{3}$P$_{0}$ states \cite{Porsev:04}.

\begin{figure}
\includegraphics[height=8.6cm,angle=270]{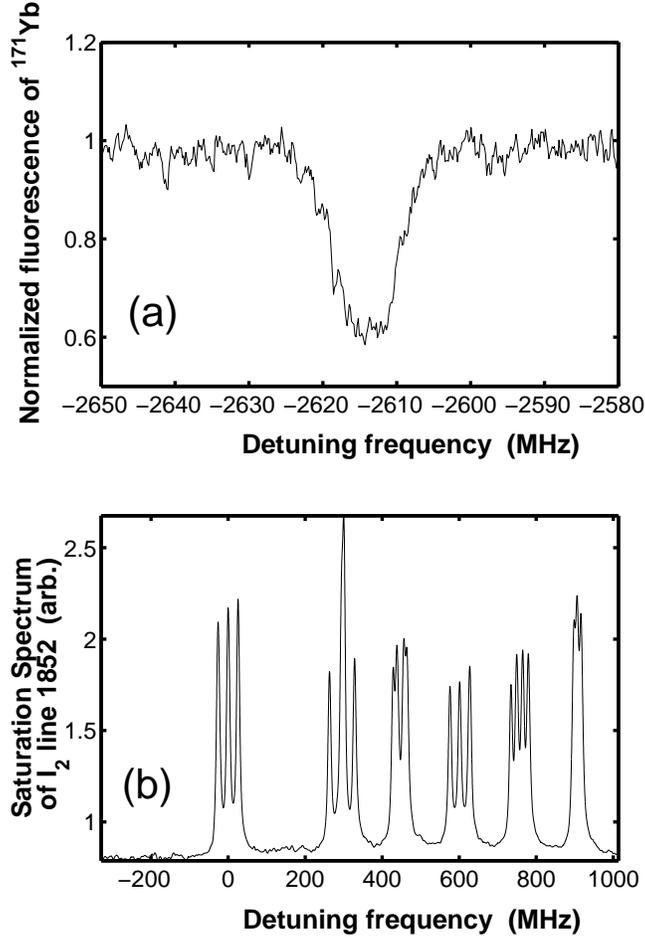}
\caption{ (a) Normalized MOT blue fluorescence of $^{171}$Yb, showing the clock transition resonance, and (b) saturation spectrum of molecule iodine reference line 1852 vs detuning frequency of the probe laser. In (a), the probe beam has a power of  0.12 W and a diameter of 4 mm.}
\label{fig:ybsat}
\end{figure}

The resonant absorption cross-section for the $\lambda = 578$ nm photons
may be written $\sigma = (\sqrt{ln2/\pi}\lambda^2/2)(\Gamma/\Delta\omega)$ for 
the case of a Gaussian line shape, where $\Delta\omega$ is the full width at half maximum, and the
natural decay rate of the $^3P_0$ state is $\Gamma \cong 5 \times 10^{-2}$
s$^{-1}$ \cite{Porsev:04}. The fractional reduction in MOT fluorescence due to the clock
transition is then $R = \sigma I/(\sigma I + \gamma_{MOT})$, where
$\gamma_{MOT}$ is the rate at which atoms are lost from the MOT in the 
absence of the clock transition.
Typically, $\gamma_{MOT} \cong 0.5$ s$^{-1}$. With our intensity
$Ihc/\lambda$ of about 1 W/cm$^2$ and a Doppler broadening of about 1MHz
at a MOT temperature of 1mK, the value of $R$ is close to unity. Thus, by contrast with a similar method in Sr \cite{French}, we
both expect and observe large fractional reduction in MOT fluorescence at the
clock resonance.

The large MOT fluorescence reduction due to the clock resonance in $^{171}$Yb
is shown in Fig.~\ref{fig:ybsat} (a). Here the resonance data are taken with the probe beam on continuously, and are subject to extra broadening of the clock
transition in the presence of the MOT light.  Data taken with the MOT and
probe beams alternately chopped show similarly large resonances with
widths comparable to the independently measured probe laser linewidth. 
Further study is required to understand and quantify the different sources
of broadening in all our resonance data. The probe laser frequency is given 
by the detuning from a specific hyperfine resonance of the iodine line 
1852\cite{IODINELINE:00}, shown in Fig.~\ref{fig:ybsat} (b).  A 
similar fluorescence reduction is also observed in the case of $^{173}$Yb. 
The two clock transitions in $^{171}$Yb and $^{173}$Yb are approximately 
2.6 and 3.9 GHz below the lowest frequency hyperfine component of the 
iodine line 1852.

Qubits, as basic elements in quantum computation, are extensively investigated in order to break through the limit of classical computers.  Decoherence of atomic qubits, often caused by external magnetic fields and collisions, is generally believed to be the largest obstacle in the way of building quantum computers.  Atom collisions can be reduced by trapping atoms in optical lattices and keeping low densities, and in fact methods for entangling qubits in lattices have been proposed \cite{Knight:03,Duan:03}. However, magnetic field perturbations can be difficult and inconvenient to eliminate in experiments. Alkaline-earth atoms show intrinsic advantages in forming qubits in this respect because their metastable $^3$P$_0$ states and the $^1$S$_0$ ground states are insensitive to external magnetic fields. These states have negligible decoherence but  reasonably large single-photon transition strengths, as shown by the large fractional fluorescence reduction even with the relatively broad linewidth in Fig.~\ref{fig:ybsat} (a).  Thus, a qubit formed by a $^1$S$_0$  ground state and a metastable $^3$P$_0$ state can simply be flipped by a single-photon transition instead of more complicated two-photon Raman transitions as for alkali atoms \cite{Knight:03,Duan:03}.   

In conclusion, we have observed the weak $^{1}$S$_{0}$-$^3$P$_0$ clock 
transition in cold $^{171,173}$Yb, which is an important step to establish Yb as one of the primary candidates for future optical frequency standards, and we have also discussed a new approach for qubits with this single-photon transition.

We would like to thank Feng-Lei Hong and Reina Maruyama for helpful
discussions. This work was supported by the National Science Foundation, Grant No. PHY 0099535.

\end{document}